\documentclass[journal]{IEEEtran}

\usepackage{subcaption}
\usepackage{tabularx}
\usepackage{diagbox}

\usepackage{xcolor}
\newtheorem{definition}{Definition}[section]

\ifCLASSINFOpdf
  \usepackage[pdftex]{graphicx}
\else
  \usepackage[dvips]{graphicx}
\fi
\usepackage{amsmath}
\usepackage{color}
\definecolor{myGreen}{RGB}{0, 102, 102}

\begin{document}
%
\title{Quantifying the Tradeoff Between Cyber-security and Location Privacy}
%
%
%

\author{Dajiang~Suo,
        M. Elena~Renda,
        and~Jinhua Zhao
\thanks{D. Suo is with the Department of Mechanical Engineering, Massachusetts Institute of Technology, Cambridge, MA, 02139 USA e-mail: {djsuo}@mit.edu.}
\thanks{M. E. Renda is with the Istituto di Informatica e Telematica, Consiglio Nazionale delle Ricerche, Pisa, Italy and MIT Department of Urban Studies and Planning, Massachusetts Institute of Technology, Cambridge, MA, 02139 e-mail: {erenda}@mit.edu, elena.renda@iit.cnr.it.}
\thanks{J. Zhao is with the MIT Department of Urban Studies and Planning,
Massachusetts Institute of Technology, Cambridge, MA, 02139 e-mail: {jinhua}@mit.edu.}}

%
%

\markboth{}%
{Shell \MakeLowercase{\textit{et Al.}}: Bare Demo of IEEEtran.cls for IEEE Journals}
%



\maketitle


\begin{abstract}
When it comes to location-based services (LBS), user privacy protection can be in conflict with security of both users and trips. While LBS providers could adopt privacy preservation mechanisms to obfuscate customer data, the accuracy of vehicle location data and trajectories is crucial for detecting anomalies, especially when machine learning methods are adopted by LBS.


This paper aims to tackle this dilemma by evaluating the tradeoff between location privacy and security in LBS. In particular, we investigate the impact of applying location data privacy-preservation techniques on the performance of two detectors, namely a Density-based spatial clustering of applications with noise (DBSCAN), and a Recurrent Neural Network (RNN).


The experimental results suggest that, by applying privacy on location data, DBSCAN is more sensitive to Laplace noise than RNN, although they achieve similar detection accuracy on the trip data without privacy preservation. 
Further experiments reveal that DBSCAN is not scalable to large size datasets containing millions of trips, because of the large number of computations needed for clustering trips. On the other hand, DBSCAN only requires less than 10$\%$ of the data used by RNN to achieve similar performance when applied to vehicle data without obfuscation, demonstrating that clustering-based methods can be easily applied to small datasets. Based on the results, we recommend usage scenarios of the two types of trajectory anomaly detectors when applying privacy preservation, by taking into account customers’ need for privacy, the size of the available vehicle trip data, and real-time constraints of the LBS application.

\end{abstract}

\begin{IEEEkeywords}
Anomaly detection, Location privacy, Differential privacy, Recurrent neural networks, DBSCAN
\end{IEEEkeywords}

%
\IEEEpeerreviewmaketitle

\section{Introduction}
%
%
%
%
\IEEEPARstart{P}{rivacy} preservation techniques have been explored and adopted by location-based systems (LBS) providers and public agencies in their data processing and analytic tasks, as location data can be used to identify an individual or to reveal customers' private information when combined with other personal identifiable information. Previous data breaches occurred in the mobility sector~\cite{uberbreach} led to concerns about the confidentiality and the potential abuse of customer data collected by LBS services. However, in cases where location-based data is aggregated for public interest, such as achieving system cyber-security, there may be a conflict between the effort to protect individual privacy and the one to ensure service security. In fact, the accuracy of vehicle location data and trajectories is crucial for determining whether trips are fabricated by adversaries or selfish drivers with malicious intent, especially when machine learning methods are adopted by LBS providers to identify trips' anomalies.

Among anomalous trips, there are the so called ``malicious trips'', whose locations have been intentionally modified by selfish drivers who want to gain economic benefits. 
Although an in-depth discussion of the reasons why these malicious behaviors occur is out of the scope of this paper, we want to point out that usually they are mainly motivated by economic profits, and they manifest themselves in different ways as technologies supporting LBS evolve in mobility sectors. As a classic example, selfish taxi drivers may intentionally prolong the routes for their customers. As a result, customers will pay higher rates for trips as  travel distances recorded by the taximeter end up to be longer than necessary with that given origins and destinations~\cite{ge2011taxi}. In the era of shared mobility, with customers, drivers and service providers relying on edge devices to communicate (e.g., smart phones), adversaries could utilize third-part malicious Apps to fake digital traces of vehicle trips in order to gain digital rewards from shared-mobility providers~\cite{uberafrica,UberChina}.  
It is interesting to see that such fraudulent behaviors often occur when a company expands its ride-hailing services in a new region or country, or it is facing challenges from competitors and decides to adopt drivers' rewarding strategies to promote its services~\cite{leng2015analysis}. Unfortunately, these strategies could result in undesired “side effects,” such as drivers who start to abuse the incentive system to earn more rewards by cheating.

Engineers in charge of developing information platforms for shared-mobility services face new challenges under both security and privacy aspects. On one hand, they need to select and apply anomaly detection methods that are most appropriate for the shared-mobility context and can achieve the best performance in detecting anomalous vehicle trips. On the other hand, since users are becoming more sensitive to privacy issues and ask for more, if any technique to preserve location privacy needs to be applied, security engineers will also need to evaluate their impact on the detector of anomalies.

This paper tackles this emerging dilemma  by evaluating the tradeoff between location privacy and security in the context of ride-hailing services.
To achieve differential privacy~\cite{andres2013geo} preservation, we use two dimensional Laplace noise to obfuscate trip data considering both each individual location (location-based perturbation), and the potential temporal correlation between locations within the same trajectory (trajectory-based perturbation).
For detecting anomalies in the trips, we use two techniques based on different mathematical principles, namely a Density-Based Spatial Clustering of Applications with Noise (DBSCAN), and a Recurrent Neural Network (RNN).
To investigate the impact of applying location data privacy-preservation techniques on the performance of the detectors, we first fed them  with the real location data and then with the obfuscated one.
The experimental results suggest that, for differential location privacy, DBSCAN is more sensitive to Laplace noise than RNN, although they achieve similar detection accuracy on the vehicle trip data without privacy protection. In fact, even with small levels of privacy noise, there is a 15$\%$ decrease in area under the curve (AUC) scores for DBSCAN, while RNN only suffers a 5$\%$ decrease with the same noise level. 


Aligning with our goal of helping the general audience gaining a better understanding of the impact of arising malicious behaviors in shared mobility, and enabling security engineers to select anomaly detectors while taking into account privacy issues, this paper provides the following contributions:

\begin{itemize}
    \item A comprehensive review of different types of adversarial behaviors in the mobility sectors and, for each of them, the motivating factors and attack methods are provided. 
    \item A comparison of the performance of two techniques for malicious trip detection. Furthermore, for each detection technique, we evaluate the impact of applying location data privacy mechanisms on their performance. The results from experimental evaluation allow us to quantify the tradeoff between location privacy and security.
    \item A set of recommendations for engineers to select anomaly detectors based on specific business needs, the size of datasets, and privacy requirements.
\end{itemize}

The paper is organized as follows: Section II gives an overview of previous work on detection methods and differential privacy, and their application to location-based systems; Section III provides motivation for conducting this research work; Section IV describes the proposed evaluation framework for quantifying the tradeoff between location privacy and security, including the vehicle trips dataset used, the obfuscation mechanism that adds two-dimensional Laplace noise to location data, the DBSCAN clustering algorithm and the recurrent neural network (RNN) for classifying vehicle trip into malicious or normal ones. In Section V, we show the experiments we conducted to test the detection performance of RNN model and DBSCAN as we vary the noise level added to trip locations/trajectories, and we provide some discussion on the results. Section VI concludes by discussing limitations of this work and potential future developments.  

\section{Previous Work}
Approaches for detecting anomalous trajectories can be categorized into rule-based and learning-based. The former has been focused on mining normal travel patterns from GPS data and conduct anomaly checking based on these patterns~\cite{zheng2015trajectory}. The basic idea is to first define clusters of ``normal trajectories'' between an origin-destination pair as routes that a majority of vehicles will take. A new trajectory can then be determined as anomalous if its driving patterns such as choice of road segments, average speed, or travel distance are significantly different from those in the normal clusters. Liu et Al.~\cite{liu2013fraud} propose a speed-based fraud detection system to detect a taxi driver’s overcharging customers. Wang et Al. conduct trajectory clustering between pickup and drop-off points and measure the similarity among trajectories for the same cluster to determine which one is anomalous~\cite{wang2018detecting}. Zhang and Chen propose an “isolation-based” detection method that can not only determine whether a route is anomalous or not, but also tell which road segments in the subsets of the route are accountable (\cite{chen2011real,chen2013iboat}). 

Deep neural networks have been gaining attention from both academia and industry to deal with GPS trajectory data for prediction tasks, such as vehicle or trajectory classification (\cite{simoncini2018vehicle,yao2018learning}), travel prediction (\cite{wang2019exploring,fan2018deep,qian2019vehicle}), and characterizing driving styles~\cite{dong2016characterizing}. For trajectory data of vehicle movement, although some research suggests that extracting semantic and statistical features can improve the performance of deep learning models (e.g., the accuracy of detection, receiver operating characteristics curve, etc.), it requires fewer efforts than rule-based approaches in features engineering. Besides, deep neural networks can deal with GPS sequences with variable length~\cite{de2015artificial}. In particular, for malicious behaviors in the context of ride-hailing, service providers have been studying the use of deep neural networks such as convolution neural networks (CNN), Long-term short Memory (LSTM), and Generative neural networks (GAN) in detecting malicious behaviors from drivers~\cite{UberChenAnomaly}. 

In addition to methods building on deep neural networks, there exist alternative machine learning algorithms for trajectory anomaly detection~\cite{belhadi2020trajectory}. For example, clustering-based approaches such as DBSCAN has been gaining popularity in recent years for trajectory anomaly detection~(\cite{belhadi2020deep,lv2017outlier}).

The notion of differential privacy is first introduced by Dwork to protect the privacy of every individual whose data is aggregated in a database for statistical analyses~\cite{dwork2008differential}. Chatzikokolakis extends standard definition of differential privacy to make it applicable to location-based services~\cite{andres2013geo}. The concept captures the level of distinguishability that a probabilistic algorithm (thereafter referred as mechanism) achieves when applying it to two adjacent datasets (two datasets are said to be adjacent if they only differ in one element)~\cite{dwork2008differential}. 

Similar ideas that adapt differential privacy to location-based crowd sensing scenarios have also been explored in~\cite{giraldo2020adversarial,wang2020sparse,wang2019location}. 

Although in all the work presented above, as well as in this work, the notion of differential location privacy is explored, the assumption of adversarial behaviors is very different. The goal of adversaries in~\cite{giraldo2020adversarial} is to cause erroneous aggregated measurements (e.g., traffic flow in a given road segment) derived from data shared by a group of vehicles. In~\cite{wang2020sparse} and~\cite{wang2019location}, the adversaries want to infer the approximate locations of users. In our work, the adversaries  are trying to fake trajectories of individual users to fool the anomaly detector.

In addition to the notion of \textit{differential location privacy}, which prevents adversaries from inferring the single location of each mobile-platform user, there exists literature that investigates a different but related notion, called \emph{differential trajectory privacy}~(\cite{chen2019real,chatzikokolakis2015geo}), which aims at avoiding the tracking of user trajectory within a period of time. Our work considers the impact of both location privacy and trajectory privacy on anomalous trip detection.

It is important to outline that the need for location privacy, e.g. obtained through data obfuscation techniques, and trajectory anomaly detection, e.g. through machine learning (ML) methods, are at odds, as the former tends to reduce the accuracy of location data which, in turn, might reduce the effectiveness of anomaly detection. The negative impact of differential privacy mechanisms on ML model accuracy is explored in~\cite{bagdasaryan2019differential}, where the authors also provide an insightful explanation of this negative impact from the perspective of gradient updates. While in~\cite{bagdasaryan2019differential} the authors focus on supervised learning algorithm, where the classification task is performed with pre-defined class labels,  our work focuses on unsupervised learning algorithms, where the anomaly detection classification is performed without any label information. The proposed evaluation framework can be used by designers and security engineers to find a tradeoff between protecting location privacy for individual customers and detecting anomalous trips with unsupervised learning techniques.

\section{Motivation}
Malicious behaviors related to location-based data started to arise among greedy drivers in traditional taxi services, where taxi drivers could play tricks to inflate charges to customers. For example, a taxi driver could manipulate the scale of the taximeter so to record a larger travel distance and faster speed than a normal one does~\cite{liu2013fraud}. Some drivers could also intentionally prolong the route for the same purpose~\cite{ge2011taxi}. 

On the security vulnerabilities respect, the integration of mobile sensing and wireless communication technologies into urban mobility systems have created new attack options, including:

\begin{itemize}
    \item Location spoofing attacks on mobile-sensing platforms. Consumer devices, such as smartphones or GPS-based navigation systems installed on vehicles, often become the target of location spoofing. An external adversary with a customized GPS spoofer could compromise the navigation system of nearby devices, with the goal to get the mobile navigation system plan incorrect routes by injecting manipulated GPS locations~\cite{zeng2018all}. An inside adversary could instead install malware or third-party Apps on a mobile device to create fake digital trajectories to be sent to the cloud server owned by mobility service provider~\cite{uberafrica}. In this case, the goal of the inside adversary could be to abuse the incentive systems provided by mobility service providers to gain digital rewards~\cite{UberChina}. For example, Uber drivers in Nigeria utilized open-source Apps to modify GPS traces to increase the travel distance and inflate charges to customers~\cite{uberafrica}. Another example involves some Uber drivers in Beijing, China, sending fake vehicle trips to the mobility service providers without physically going to the claimed locations, since whoever finishes a certain number of trips will be rewarded a certain amount of points to be cashed later~\cite{UberChina}. 
    These location-spoofing attacks are quite similar to the ones occurred in virtual games built upon mobile sensing platforms, such as Pokémon Go~\cite{zhao2017location}.
    \item Location spoofing attacks on Vehicle-to-Vehicle (V2V) communications. The adoption of V2V communication could create security vulnerabilities that could be utilized by inside adversaries to launch cyber attacks. For example, a compromised vehicle might broadcast malicious location messages to influence the driving maneuver of nearby vehicles~\cite{so2019physical}. 
    \item Location spoofing attacks on Vehicle-to-Infrastructure (V2I) communications. Similar to V2V communication, V2I communication could also suffer cyber threats resulting from location spoofing. A vehicle with malicious intent could send  through V2I channels false trajectory information to the road infrastructure, such as, for instance, a traffic signal controller~\cite{chen2018exposing}. Traffic controllers use vehicle shared location and trajectory information to optimize signal timing, and in presence of false trajectories would follow an non-optimal strategy for adjusting signal phases. This will create extra intersection delays and increase the chance of traffic congestion.
\end{itemize}

To mitigate cyber threats due to location spoofing, mobility services and infrastructure providers have developed analytical and data-driven models built from regular vehicle trips to detect malicious trips containing anomalous trajectories~(\cite{chen2013iboat,oh2019sequential,gray2018coupled}). The effectiveness of a model for anomaly detection in LBS depends not only on the sophistication of the model but also, and more importantly, on the accuracy and correctness of collected location data. In emerging mobility systems, though, the collection, sharing, usage, and storage of location data  also raise customer privacy concerns, since location data could reveal valuable information on customer behaviors/habits. In order to gain customer trust and meet privacy laws and regulations, like the EU General Data Protection Regulation (GDPR)~\cite{goddard2017eu} and the California Consumer Privacy Act~\cite{harding2019understanding}, mobility service providers may choose to apply privacy preservation techniques to vehicle trips, such as obfuscating original location data, during the data collection process. The problem is that the noise introduced to protect location data might affect the effectiveness of models built to detect for malicious behaviors. 

 To show how noise can be injected into location-based data in mobile sensing platforms to achieve some privacy level, here we present three typical architectures that have been explored for privacy preservation, namely the client-server (centralized), the distributed, and the trusted third-party location server architecture~\cite{chow2009privacy}.

\begin{figure*}[tb!]
	\centering
	\begin{subfigure}{0.45\textwidth}
		\centering
		\includegraphics[width=\textwidth]{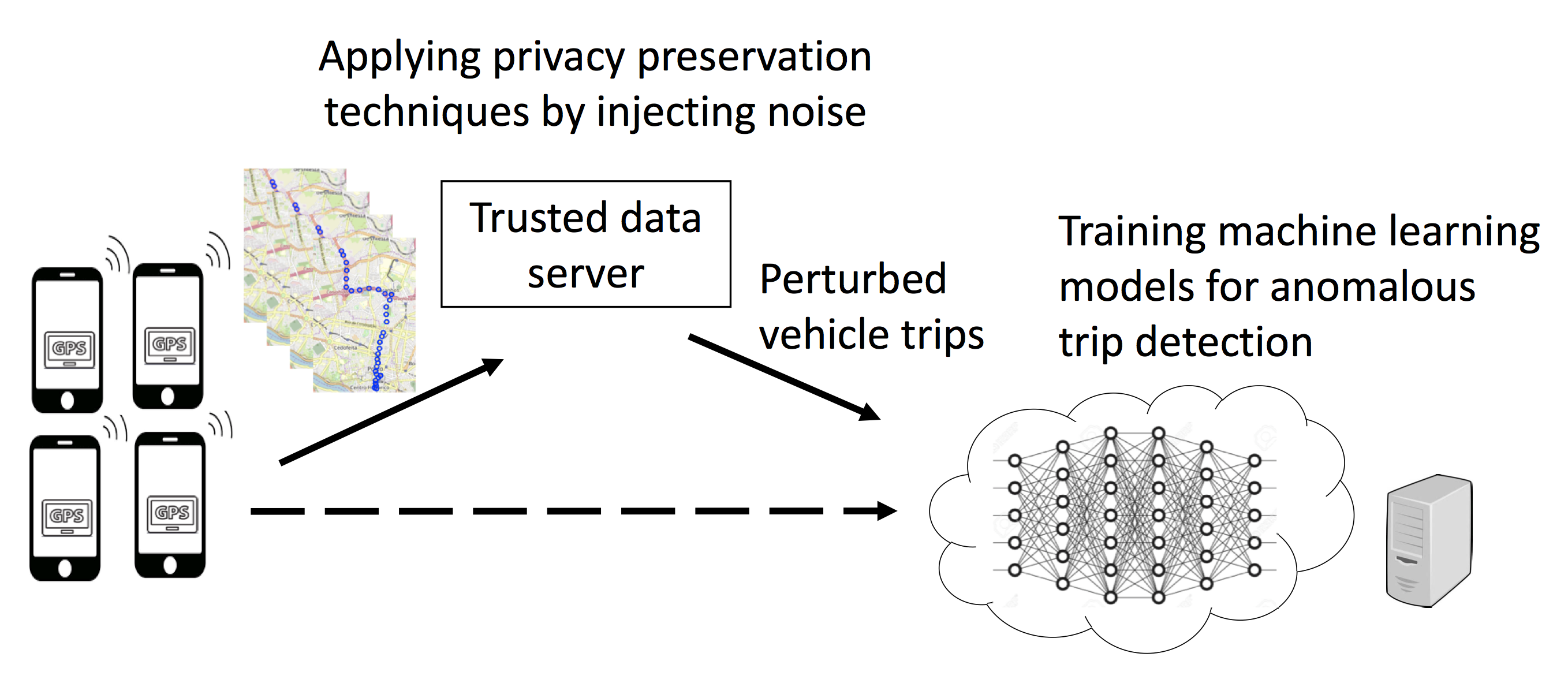}
		\caption{Training stage: GPS-generated trajectories are obfuscated by adding two-dimensional noise, before being sent to the cloud server owned by the mobility service provider, in order to train the ML model.}
		\label{fig1a}
	\end{subfigure}
	\begin{subfigure}{0.45\textwidth}
		\centering
		\includegraphics[width=\textwidth]{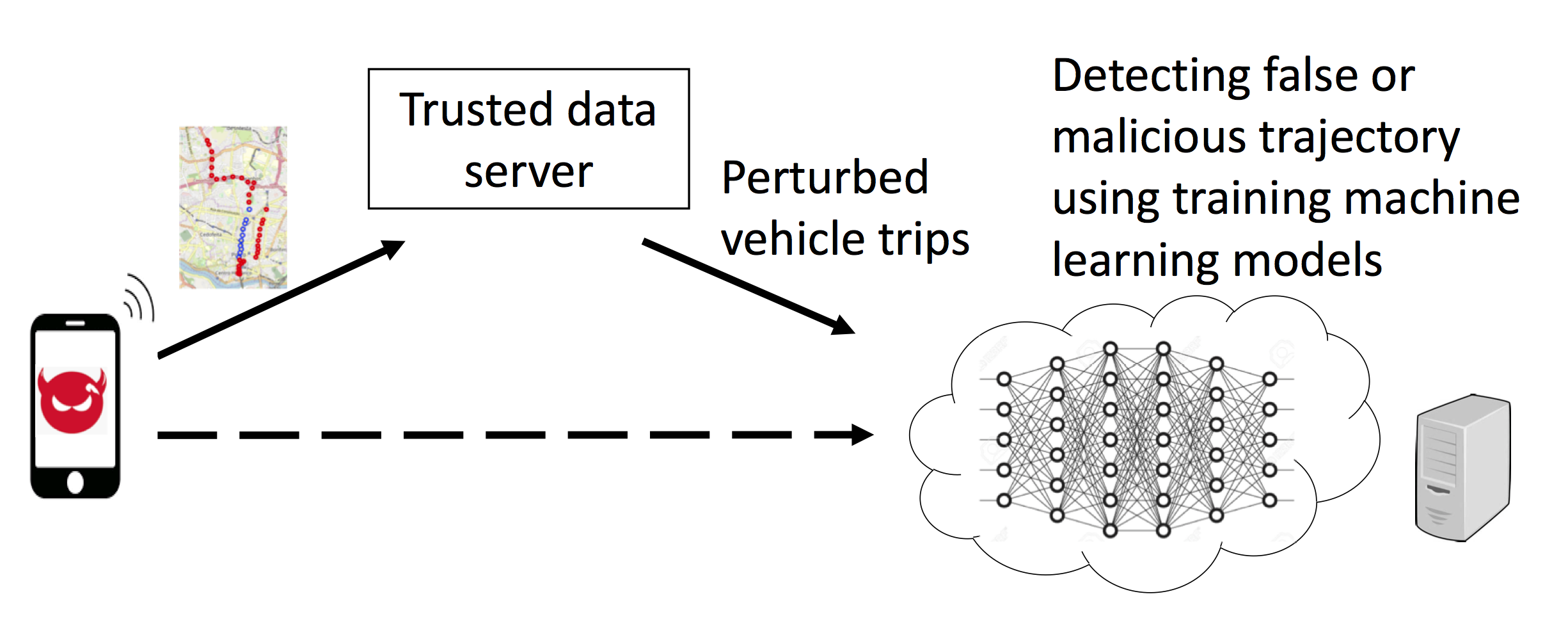}
		\caption{Detection stage: The trained ML model can determine if a vehicle trip was tampered or faked by adversaries.}
		\label{fig1b}
	\end{subfigure}
	\caption{The proposed framework for quantifying the tradeoff between location privacy and security. To understand this tradeoff, we quantify the influence of different levels of noise (i.e., level of privacy protection) added to location data on the detection accuracy.}
	\label{fig1}
\end{figure*}

\begin{itemize}
    \item In the client-server architecture~\cite{hoh2007preserving}, each vehicle sends its GPS traces to a secure and trusted data server within a centralized fleet management and traffic control center. The data server applies privacy-preservation techniques to the GPS traces before permanently storing and/or sharing them for analytic tasks.
    \item In the distributed architecture, the algorithm for privacy preservation is deployed within the users’ edge device~\cite{elmisery2019new}, hence location data will be obfuscated or sampled before being sent. 
    \item In the trusted data server architecture~(\cite{ma2019real,xu2018distilling}), location data can only be accessed through secure APIs, even for requests coming from the location service provider. This design has the advantage of protecting location data against data breaches by inside attackers, at the expense though of increased deployment cost and communication overhead between the location data server and the LBS. Every vehicle periodically updates its status by sending location-related data to the server. The LBS delegates the tasks of processing location data, such as route scheduling and path planning, to the location data server every time a new request is received from vehicles. If there is the need to publish these datasets or send data to the LBS, the server applies data preservation techniques beforehand.
\end{itemize}

While the client-server architecture can protect customer data from external exploitation of publicly available datasets, the last two architectures aim at protecting data confidentiality against inside adversaries in the event of data breaches. 

This paper focuses on privacy risks arising from inside adversaries, and we assume that either the customer mobile sensing device or a trusted third-party data server will apply data preservation techniques to the collected location data, before the data is sent to the mobility service provider for any analytic tasks. The proposed framework for anomalous trip detection in presence of privacy preservation and the evaluation of the privacy-security tradeoff is shown in Fig.~\ref{fig1}.

\section{Methodology}
This paper explores the tradeoff between location privacy and security in ride-hailing services by evaluating the impact of differential location privacy schemes on different anomalous trip detectors. As shown in Fig.~\ref{fig1}, without privacy preservation trajectory information of each vehicle trip will be directly sent (dashed lines) from the customer edge device to the mobility service provider, which relies on machine learning models to determine if the trip was tampered or faked. However, to meet customers' need for privacy, the edge device will be required to send trip data to a trusted server in charge of obfuscating GPS locations before routing them to the mobility service provider. 

Since adding noise to location data can result in a degraded performance in detecting anomalies, we will try to understand how and to what extent differential location privacy will decrease the accuracy of the detectors, by comparing the performance they achieve with the original data and the one they achieve with the obfuscated data (i.e., without and with privacy).

In the following, after some preliminaries, we describe in detail the malicious trip generation phase, the location perturbation phase, and the malicious trip detection phase.

\subsection{Preliminaries}\label{sec:Preliminaries}
\begin{definition} A trajectory $X_n$ is a sequence of tuple ${x_1,...,x_n}$ where $x_i$ represents the geo-location coordinates $(log_i,lat_i)$ of a vehicle at a certain time step. A mobility service provider will use the trajectory data to determine if the reported trip is malicious. Here, we assume that the mobility service provider will have to make queries through the trust data server, rather than getting direct access to the trajectory data.

\end{definition}

\begin{definition}\label{def:geo}
Geo-indistinguishability~\cite{andres2013geo}.
Assume $K:X \rightarrow P(Z)$ to be a probabilistic mechanism that maps each location element in $X(x \in X)$ to a location in $Z(z \in Z)$ with a probability $P(Z)$. Then, a mechanism $K$ is said to be $\epsilon -geodistinguishable$ if and only if for all $x,x'$, we have
\end{definition}

\begin{equation}\label{eq:geodisting}
d_p(K(x),K(x')) \leq \epsilon d(x,x') 
\end{equation}

where $d_p(.,.)=sup_{x,x'\in X}\left|\frac{K(x)}{K(x')}\right|$ represents the multiplicative distance between two distributions; and $d(x,x')$  denotes the Euclidean distance between two locations $x$ and $x'$. Moreover, Equation~(\ref{eq:geodisting}) can be re-written as $K(x) \leq e^{\epsilon d(x,x')K(x')}$, where $\epsilon$ denotes the inverse of the added noise. Intuitively, in an LBS adopting mechanism $K$ for privacy preservation, the user experiences $\epsilon d(x,x')$ privacy, because the probability that the mechanism reports the location $z$, given input locations $x$ and $x'$, only differs by a multiplicative factor. In other words, the smaller is the value for $\epsilon$, the more noise will be added to the original locations.

\begin{figure*}[tb!]
	\centering
	\begin{subfigure}{0.44\textwidth}
		\centering
		\includegraphics[width=\textwidth]{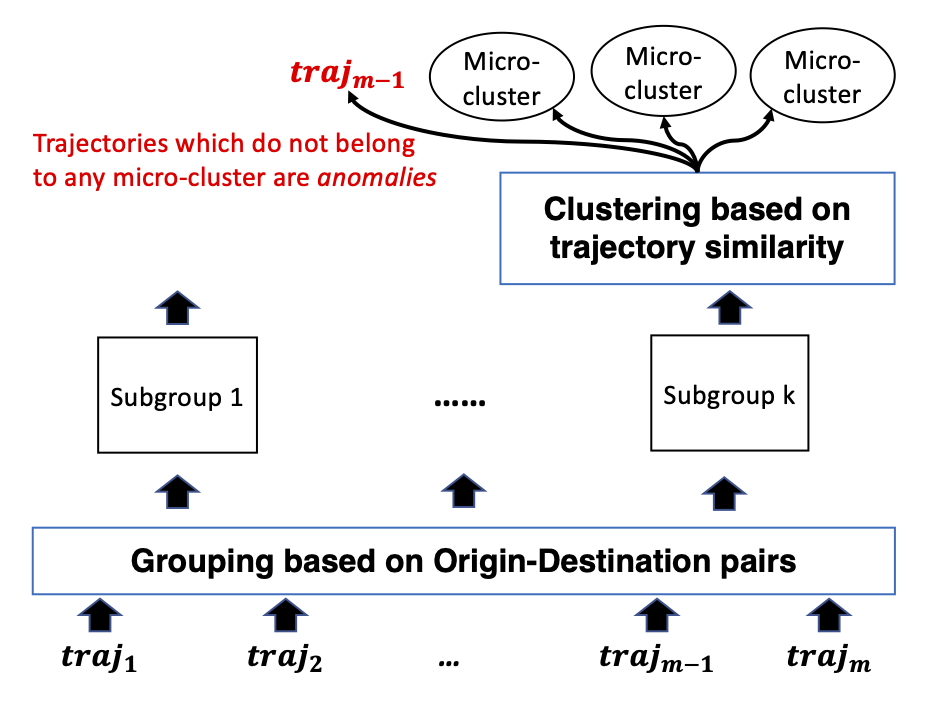}
		\caption{DBSCAN architecture for anomalous trip detection.}
		\label{fig2a}
	\end{subfigure}
	\begin{subfigure}{0.50\textwidth}
		\centering
		\includegraphics[width=\textwidth]{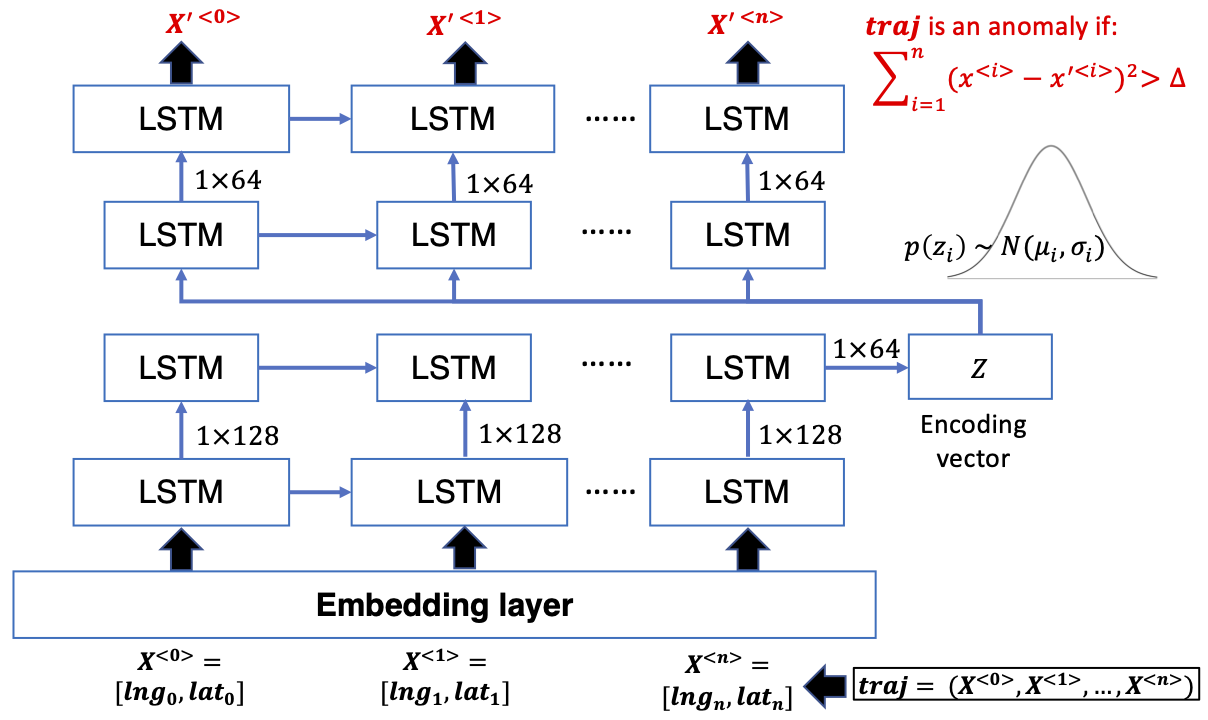}
		\caption{RNN architecture for anomalous trip detection~\cite{liu2020online}.}
		\label{fig2b}
	\end{subfigure}
	\caption{Anomalous trip detectors built with two different unsupervised learning techniques.}
	\label{fig2}
\end{figure*}

\subsection{Attack models and malicious trip generation}
To evaluate the impact of privacy-preservation mechanisms on anomalous trip detectors, we need a groundtruth dataset containing both normal and malicious trips, labeled. However, no such publicly available dataset exists to the best of our knowledge, so we had to generate one. There are three common strategies for deriving anomalous trajectories from normal vehicle trips, to later use for model evaluation. The first one is to inject trip trajectories that do not belong to any trip used in training, as suggested by Min-hwan~\cite{oh2019sequential}. The second one involves extracting and isolating special trips that significantly differ from the rest of the training dataset and define them as anomalous. For example, Gray et Al. isolate trajectory data of different transportation modes (private car vs. bike or public transit) and treat the former as malicious~\cite{gray2018coupled}. The third type of strategy is to formulate the anomalous trip generation as an optimization problem with an explicit attack goal and geospatial constraints~\cite{huang2020impact}. We adopt an approach that merges the first and third strategies. The trips dataset we use in our experiments is the Portugal taxi trajectory dataset~\cite{portodata}, which includes trajectories from 442 taxis in Porto, Portugal between July 1st, 2013 and June 3rd, 2014. It was released in Kaggle ECML/PKDD 15: Taxi Trajectory Prediction (I) competition, and it has been widely used in academia for benchmarking new algorithms for trajectory modeling~\cite{qian2019vehicle,de2015artificial,besse2017destination} and malicious trip detection~\cite{oh2019sequential}.
We generate and inject new (malicious) trips into the Porto dataset considering simplified assumptions on attacker's goals and constraints, as shown in the optimization problem below.

\begin{subequations}\label{eq:opt}
\begin{align}
\max_{X_{n'}} \quad & ||R(X_n)-R(X_{n'})||_2 \label{eq:opta}\\
\textrm{s.t.} \quad & sim(X_n,X_{n'}) \leq B\label{eq:optb}\\
  & m/n \leq q   \label{eq:optc}\\
  & x_{n,i} \in X_n, x_{n',i} \in X_{n'} \label{eq:optd}\\
  & X_{n'} \in \Omega_{x_{map}}\label{eq:opte}
\end{align}
\end{subequations}

The objective function in Equation~\ref{eq:opta} represents the adversary's goal of maximizing the difference between the rewards deriving from a normal trajectory $X_n$ and the tampered/fabricated trajectory $X_{n'}$. Equation~\ref{eq:optb} sets a constraint on the shape of the whole fabricated trajectory by the adversary. The similarity function $sim(X_n,X_{n'})$ measures the difference between $X_n$ and $X_{n'}$ and $B$ denotes the total budget for trajectory differences~\cite{wang2013effectiveness}. In its simplest form, the similarity function can be expressed as $d(x_{n,i},x_{n',i})=c$, where $c$ represents the extent to which each location in the spoofed GPS trajectory diverges from the original one. Therefore, it sets a constraint on the difference between two locations in the normal and malicious trajectories respectively, meaning that the Euclidean distance between location $x_n$ in the normal trajectory and location $x_{n'}$ in the malicious trajectory must be less than distance $c$ (meters). Equation~\ref{eq:optc} denotes the proportion of location points that are tampered by an adversary, where $m$ represents the number of tampered locations. Given the definitions above, we can now use the pair $(c,q)$ to describe the level of malicious intent of the adversary.

The last condition Equation~\ref{eq:opte} means that each point in the trajectory $X_{n'}$ must belong to a feasible region $\Omega$ on the map, so the whole trajectory is more plausible. For example, a location with GPS coordinates in a river or extreme altitude might be implausible~\cite{UberChenAnomaly}.

\subsection{Location perturbation for differential privacy}

For perturbing vehicle location data to achieve differential location privacy, we adopt Geo-Indistinguishability~\cite{andres2013geo} by adding two dimensional Laplace noise to the trip data. Definition~\ref{def:geo} provided in Section~\ref{sec:Preliminaries} gives insights on how an edge device or a trusted data server, given the actual location $x$ of the user, could obfuscate $x$ to generate the location $x'$ before sending it to the mobility service provider. The function for achieving this purpose is $D_{\epsilon}(x)(z)=\frac{\epsilon^2}{2 \pi}e^{-\epsilon d(x,z)}$, which follows a two-dimensional Laplace distribution. Intuitively, this distribution means that the probability that the mechanism $K$ generates a point near $x$ decreases exponentially as the Euclidean distance between $x$ and $x'$ increases.

In addition to the Geo-Indistinguishability mechanism, which perturbs individual locations within each trajectory, we also investigate how the correlation among locations could influence the perturbation algorithm. For scenarios involving real-time (and continuous) reporting of the user location, an adversary who has gained some prior knowledge from previous locations reported by the user could utilize Bayesian inference techniques to predict the next possible locations the user may visit, even if all previously reported locations have been perturbed~\cite{xiao2015protecting}. For example, for a vehicle traveling from the west to the east, if Geo-Indistinguishability happens to shift the first five user-reported locations all to the north, then it will not be wise to perturb the next (i.e., sixth) location to a point locate in the south. This is because the adversary may gain the knowledge that the approximate region the user may visit next is more likely to locate north to the horizontal line formed by the first five locations.

To resolve the concern arising from temporal correlation between locations, we adopt an algorithm called Predicative Mechanism, derived from~\cite{chatzikokolakis2015geo}, and use it as a benchmark for Geo-Indistinguishability in our experiments. For the sake of brevity, we refer to Geo-Indistinguishability as \emph{location-based perturbation} and the privacy preservation mechanism considering temporal correlation~\cite{chatzikokolakis2015geo} as \emph{trajectory-based perturbation}.

\subsection{Malicious trip detection}
This paper evaluates both RNN and DBSCAN for anomaly detection in vehicle trajectory data while applying location privacy. These two models have been selected as each one is representative of one category of techniques often used in trajectory anomaly detection: non-parametric distance-based clustering (DBSCAN)~\cite{ester1996density} and pattern mining based on recurrent neural networks (RNN)~\cite{belhadi2020trajectory}. From the perspective of data processing, there are two major differences between DBSCAN and RNN. First, unlike RNN, DBSCAN can be applied directly to vehicle trip data for detecting anomalies without training, which often requires the collection of a large amount of vehicle trip data. Second, DBSCAN assumes that the information of complete vehicle trajectories is available (i.e., off-line detection), especially origins and destinations, while RNN can be applied to vehicle trips that only contains partial trajectory information. For this reason, RNN is often used in on-line anomaly detection when mobility service providers need to collect each vehicle's locations in real-time. To ensure a fair comparison between DBSCAN and RNN, we use vehicle trip data with complete trajectory information, as required by DBSCAN. In the following we introduce both techniques.

\subsubsection{DBSCAN}
DBSCAN has been used in literature for identifying outliers from trajectory datasets~(\cite{belhadi2020deep,lv2017outlier}). As shown in Fig.~\ref{fig2a}, the key idea is to group a set of vehicle trips into different clusters and find trips that do not belong to any cluster. To achieve this goal, a metric that measures how similar any two trajectories are must be defined. This paper extends the DBSCAN algorithm derived in~\cite{lv2017outlier} by adopting a new metric, called Frechet Distance~\cite{alt1995computing}, to measure the similarity between different trajectories. Specifically, given two curves (i.e., trajectories in the context of our discussion) $A$ and $B$ each of which is a mapping from $[a,b]$ ($[a',b']$) onto a metric space $S$, their Frechet Distance is defined as 

\begin{equation}\label{eq:frechet_Dist}
F(A,B) = \underset{\alpha,\beta}{inf}\underset{t\in[0,1]}{max}\{d(A(\alpha(t)),B(\beta(t)))\}
\end{equation}

where $\alpha$ and $\beta$ are arbitrary continuous non-decreasing functions mapping from $[0,1]$ to $[a, b]$ and $[a',b']$ respectively. In our implementation of DBSCAN, we use the discrete version of Frechet Distance ~\cite{eiter1994computing} since each trip consists of discrete geo-locations sampled from GPS devices.

A trip belonging to a given cluster has smaller ``distance", in terms of similarity, from trips within the same cluster, than from trips belonging to a different cluster.

\subsubsection{RNN}
We are also interested in an “end-to-end” malicious behavior detection method~\cite{gray2018coupled} by using deep neural networks. Unlike DBSCAN, which requires the manual selection of the minimum number of trips for each cluster, this type of method takes original trip data and output a label indicating whether the trip is malicious or not. Specifically, we adopt RNNs, a specialized type of neural network that is good at processing sequential data (e.g., GPS trajectory)~\cite{pineda1987generalization}. It takes inputs at every (distributed) time step and gives outputs at either each step corresponding to inputs, or at the last step. In our analyses, we use the Long-Short Term Memory (LSTM) model~\cite{hochreiter1997long}, which is a modified version of RNNs. The reason is that the former is capable of learning long-term spatial dependencies between GPS points that are far apart from each other~\cite{qian2019vehicle}, namely, those GPS points that are more likely to co-occur in vehicle trajectories. 
The actual architecture of RNNs we use as the benchmark for testing privacy preservation techniques in this paper is called "Gaussian Mixture Variational
Sequence AutoEncoder" (GMVSAE)~\cite{liu2020online} (shown in Fig.~\ref{fig2b}), which achieved high performance in anomaly detection for the Porto dataset  we use for our experiments. In the training stage, given GPS trajectory $X$ as input, the encoder part learns (low-dimensional) latent vectors $Z$, which is used to reconstruct the inputs as $X'$. In the prediction stage, any new trip that got a high reconstruction error (denoted as $\frac{1}{n}\sum^n_i(x_i-x_i^{'})^2$) will be regarded as malicious. One feature of the GMVSAE is that the latent vector is probabilistic and it is enforced to follow Gaussian distribution ($p(z)$ in Fig.~\ref{fig2b}). It is worth noting that, despite the selection of the specific neural network architecture in this paper, the framework for privacy-security tradeoff evaluation is applicable to other scenarios where machine learning algorithms are used for detecting anonymous behaviors.

\section{EXPERIMENTAL EVALUATION}
\subsection{EXPERIMENTAL SETUP}
As mentioned earlier, in our experiments we use the Portugal taxi trajectory dataset~\cite{portodata}.

Since we would like to choose the state-of-the-art deep neural network as the benchmark for comparing the performance of the proposed clustering-based approach, we use the same setting for the Porto taxi dataset used in~\cite{liu2020online}, which achieves the best performance in trajectory anomaly detection by using RNNs. Specifically, we select 251,550 trips which do not have gps errors and have a higher gps sampling rate (more than 25 gps points in a trajectory) and group these trips into 428419 groups each of which has a unique origin-destination (OD) pair. For each OD group, 5 trips are randomly selected for testing while the rest are used for training, leading to a total of 11,811 testing trips. 

Clearly, the same trips are used as input to DBSCAN, since it can be directly applied to anomalous trip detection without training. It is worth mentioning that, unlike the RNN-based approach, DBSCAN requires the manual selection of parameters (i.e., the minimum number of trips within each micro-cluster, as shown in Fig.~\ref{fig2a}) every time a new Laplace noise level is chosen for privacy preservation. In our experiments, the value of the minimum number of trips within each micro-cluster ranges between 2 and 5 for three levels of $epsilon$.

\subsection{Simulation results and discussion}
In this paper we want to evaluate the impact of different privacy perturbation techniques on the two classifiers adopted, i.e., DBSCAN and RNN. For each classifier, we apply both location-based and trajectory-based perturbation to vehicle trip data before feeding the data to the classifier for detecting anomalies. Additionally, we conduct experiments to understand to what extent shifting origin-destination pairs in each trajectory could influence the performance of the classifiers. 

To quantify the impact of perturbing trip location data on the classifiers' performance, we adopt receiver operating characteristics curve (ROC) and AUC scores as the performance metrics in detecting anomalous trips. Also, we control and change the values of two parameters during the experiments, namely the level of two-dimensional Laplace noise ($\epsilon$) added to the data, and the level of malicious intent (denoted as $(c,p)$ according to the definition provided following Equation~\ref{eq:opt}. By varying these two parameters, engineers can evaluate the performance of DBSCAN and RNN and decide whether to increase or decrease the privacy level (e.g., Laplace noise) to satisfy customers with different privacy needs, or based on the behaviors of adversaries. 

\subsubsection{The influence of privacy perturbation}
\begin{figure*}[tb!]
	\centering
	\begin{subfigure}{0.3\textwidth}
		\centering
		\includegraphics[width=\textwidth]{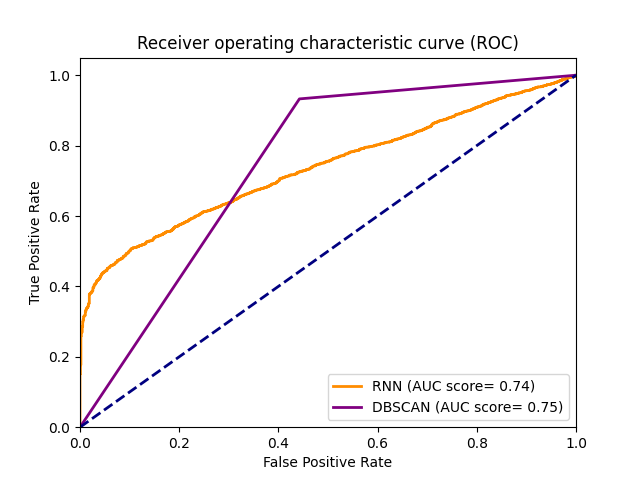}
		\caption{Without privacy protection}
		\label{fig:mal300_nopri}
	\end{subfigure}
	\begin{subfigure}{0.3\textwidth}
		\centering
		\includegraphics[width=\textwidth]{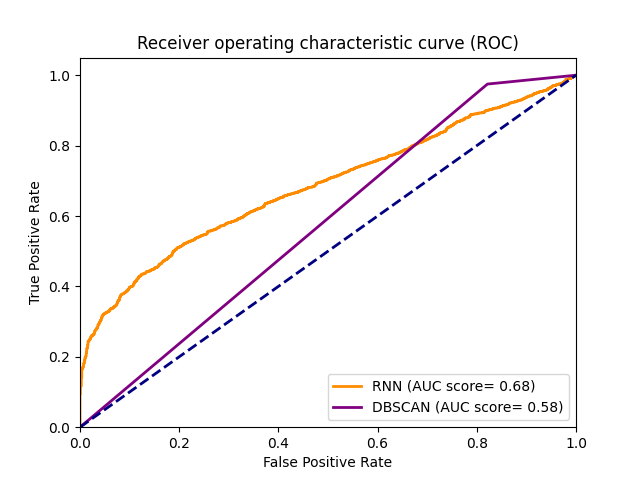}
		\caption{When $\epsilon$=0.1}
		\label{fig:mal300_eps1}
	\end{subfigure}
	\begin{subfigure}{0.3\textwidth}
		\centering
		\includegraphics[width=\textwidth]{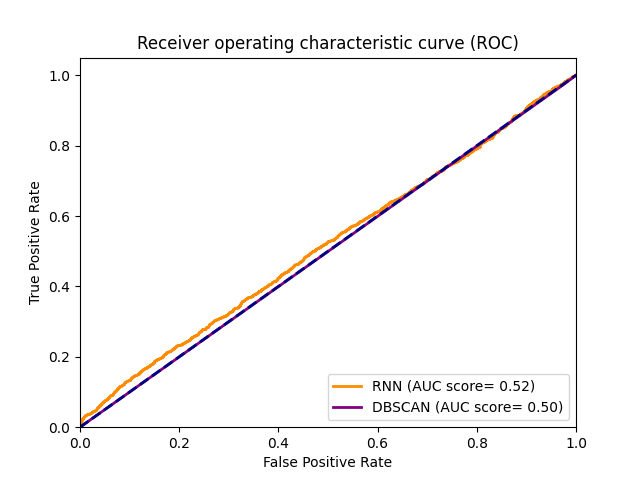}
		\caption{When $\epsilon$=0.01}
		\label{fig:mal300_eps01}
	\end{subfigure}
	\caption{Comparing the classification accuracy between DBSCAN and RNN under different differential privacy levels. The value of malicious intent is set to $(c,q)$=(300m,0.5).}
	\label{fig:twoclassifier_mal300m}
\end{figure*}

\begin{figure*}[tb!]
	\centering
	\begin{subfigure}{0.3\textwidth}
		\centering
		\includegraphics[width=\textwidth]{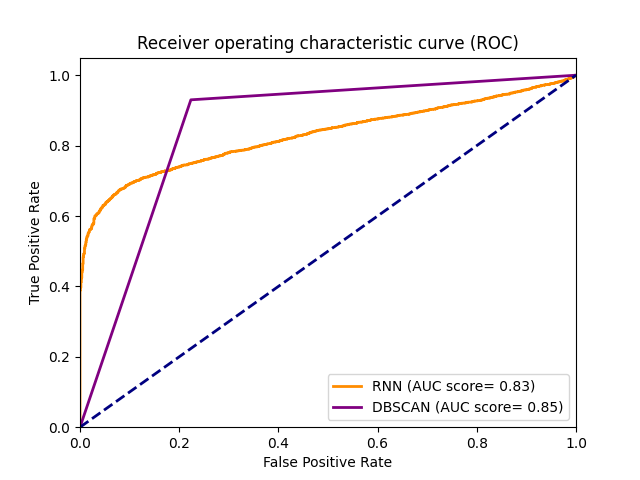}
		\caption{Without privacy protection}
		\label{fig:mal500_nopri}
	\end{subfigure}
	\begin{subfigure}{0.3\textwidth}
		\centering
		\includegraphics[width=\textwidth]{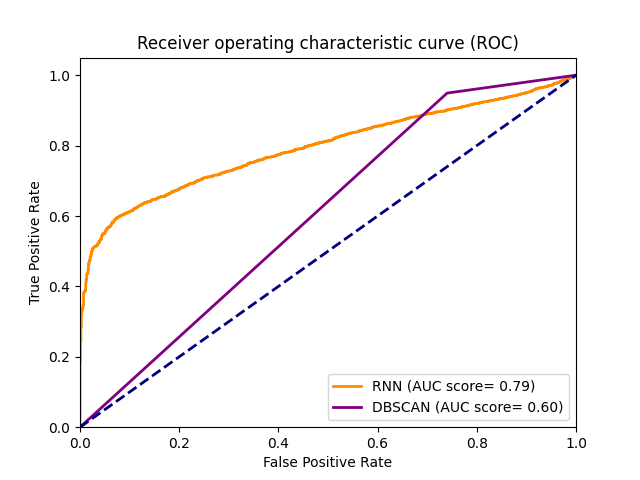}
		\caption{When $\epsilon$=0.1}
		\label{fig:mal500_eps1}
	\end{subfigure}
	\begin{subfigure}{0.3\textwidth}
		\centering
		\includegraphics[width=\textwidth]{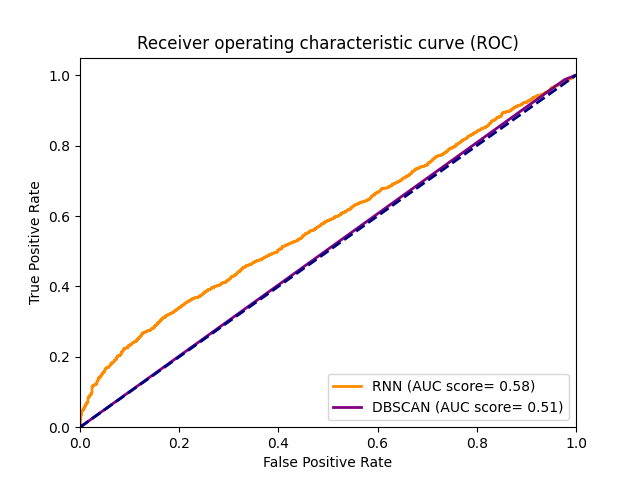}
		\caption{When $\epsilon$=0.01}
		\label{fig:mal500_eps01}
	\end{subfigure}
	\caption{Comparing the classification accuracy between DBSCAN and RNN under different differential privacy levels. The value of malicious intent is set to $(c,q)$=(500m,0.7).}
	\label{fig:twoclassifier_mal500m}
\end{figure*}

\begin{figure*}[tb!]
	\centering
	\begin{subfigure}{0.3\textwidth}
		\centering
		\includegraphics[width=\textwidth]{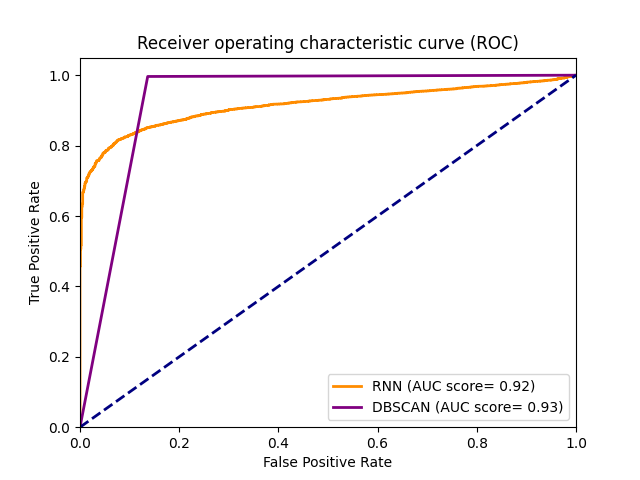}
		\caption{Without privacy protection}
		\label{fig:mal700_nopri}
	\end{subfigure}
	\begin{subfigure}{0.3\textwidth}
		\centering
		\includegraphics[width=\textwidth]{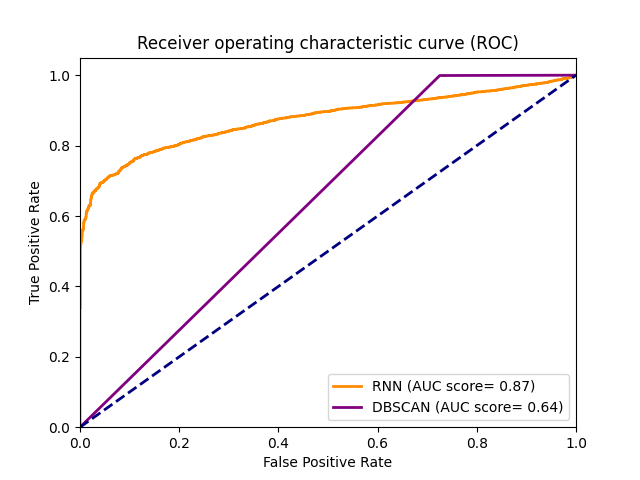}
		\caption{When $\epsilon$=0.1}
		\label{fig:mal700_eps1}
	\end{subfigure}
	\begin{subfigure}{0.3\textwidth}
		\centering
		\includegraphics[width=\textwidth]{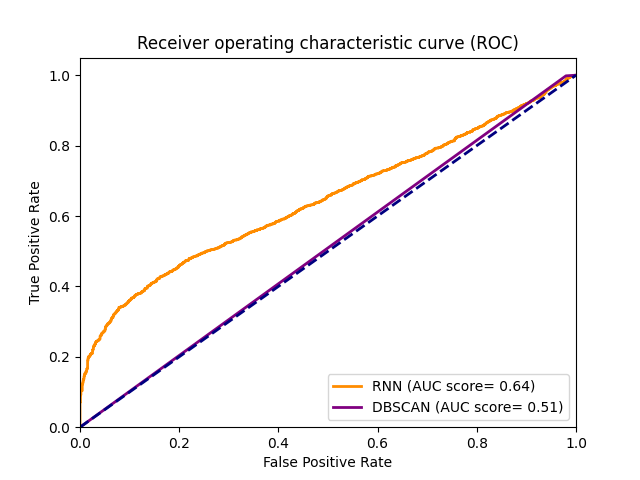}
		\caption{When $\epsilon$=0.01}
		\label{fig:mal700_eps01}
	\end{subfigure}
	\caption{Comparing the classification accuracy between DBSCAN and RNN under different differential privacy levels. The value of malicious intent is set to $(c,q)$=(700m,1.0).}
	\label{fig:twoclassifier_mal700m}
\end{figure*}

When the original vehicle locations are directly used for anomaly detection, i.e. without applying any privacy perturbation, DBSCAN and RNN achieve similar performance in detecting anomalous trips, as shown in Figures~\ref{fig:mal300_nopri},~\ref{fig:mal500_nopri}, and~\ref{fig:mal700_nopri}. As expected, by increasing privacy level through adding Laplace noise to vehicle trips results in a decreased performance in differentiating between anomalous and regular trips, for both DBSCAN and RNN. However, DBSCAN suffers a more serious performance degradation than RNN as we add more noises to the trip data. The AUC score achieved by DBSCAN drops to less than 0.65 even for small privacy perturbations ($\epsilon$ = 0.1), as shown in Figures~\ref{fig:mal300_eps1},~\ref{fig:mal500_eps1}, and~\ref{fig:mal700_eps1}. For the same level of privacy perturbation, RNN can still maintain an AUC score 10\% higher than DBSCAN's. The experimental results suggest that RNN is a better fit for situations where the collection of raw location-based data is not allowed.

In addition to location-based perturbation, we also investigate the impact of applying trajectory-based perturbation on anomaly detection. The trajectory-based perturbation is a privacy-preservation mechanism designed for perturbing vehicle trajectory whenever a new location comes in, which is often used in scenarios where on-line monitoring or continuous reporting of user locations is necessary. Since DBSCAN is an off-line anomaly detection mechanism and can only be applied when every location within a trajectory is collected, we only evaluate RNN performance with trajectory-based perturbation. 

\begin{figure*}[tb!]
	\centering
	\begin{subfigure}{0.3\textwidth}
		\centering
		\includegraphics[width=\textwidth]{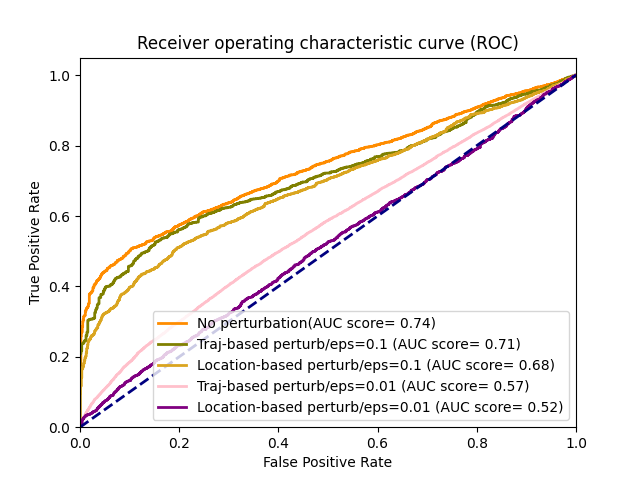}
		\caption{$(c,q)$=(300m,0.5)}
		\label{fig:rnn_mal300}
	\end{subfigure}
	\begin{subfigure}{0.3\textwidth}
		\centering
		\includegraphics[width=\textwidth]{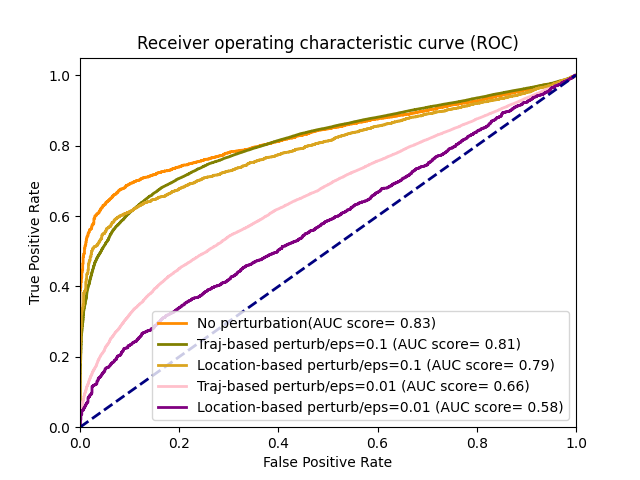}
		\caption{$(c,q)$=(500m,0.7)}
		\label{fig:rnn_mal500}
	\end{subfigure}
	\begin{subfigure}{0.3\textwidth}
		\centering
		\includegraphics[width=\textwidth]{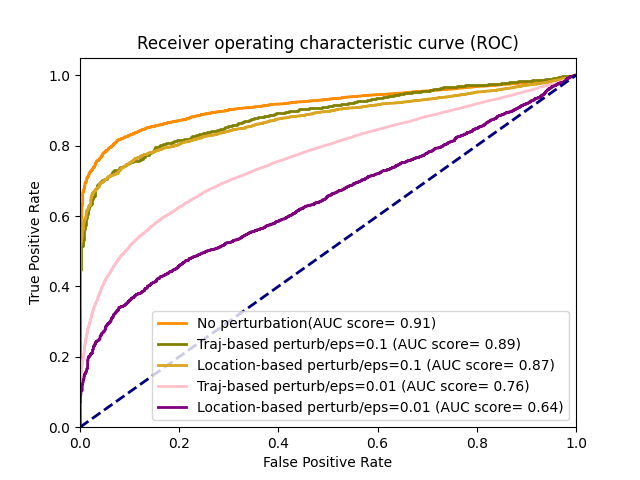}
		\caption{$(c,q)$=(700m,1.0)}
		\label{fig:rnn_mal700}
	\end{subfigure}
	\caption{The influence of considering correlations among locations within each trajectory in location perturbation on the classification accuracy by RNN. For same values of malicious intent, applying trajectory-based perturbation allows the proposed RNN model to achieve better performance in detecting anomalies than location-based perturbation.}
	\label{fig:rnn_traj}
\end{figure*}

The performance of RNN in detecting anomalous trips under different privacy perturbation mechanisms is shown in Fig.~\ref{fig:rnn_traj}. According to the simulation results, RNN achieves better AUC score when trajectory-based perturbation is applied to vehicle trip data. However, the difference between trajectory-based and location-based privacy perturbation is not significant for all malicious intent level shown in Figures~\ref{fig:rnn_mal300},~\ref{fig:rnn_mal500},~\ref{fig:rnn_mal700} (only 0.2 difference in the AUC score is observed).

\begin{table*}[!htb]
\centering
	\caption{The influence of privacy preservation techniques on anomaly detectors.}
		\begin{tabularx}{\textwidth}{|X|X|X|}
		    \hline
			\diagbox{\textbf{Privacy preservation}}{\textbf{Anomaly detection }}&\textbf{DBSCAN}&\textbf{RNN} \\
			\hline
			Location-based perturbation (i.e., Geo-Indistinguishability)&The DBSCAN classifier is extremely sensitive to location-based perturbation. The AUC score dropped to less than 0.65 even for small privacy perturbations with $\epsilon$ = 0.1. & The RNN classifier is less sensitive to location-based perturbation than DBSCAN. However, the AUC score also dropped to less than 0.65 for all malicious intent values.  \\
			\hline
			Trajectory-based perturbation (i.e., Geo-Indistinguishability considering the correlations among locations within each trajectory)& Not applicable as DBSCAN only supports off-line detection with complete trajectory information&Trajectory-based perturbation enables the RNN classifier to achieve better AUC score, especially when the privacy level is high (e.g., $\epsilon>$0.1)\\
			\hline
		 \end{tabularx}
		\label{tab:results}
\end{table*}

\subsubsection{The influence of O-D locations in generated anomalous trips}
\begin{figure*}[tb!]
	\centering
	\begin{subfigure}{0.3\textwidth}
		\centering
		\includegraphics[width=\textwidth]{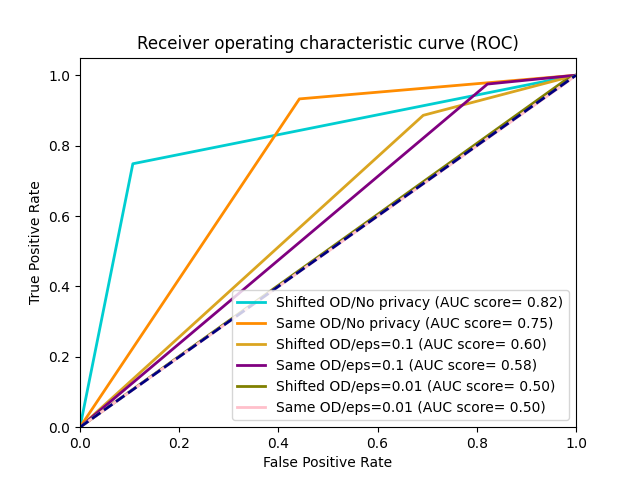}
		\caption{$(c,q)$=(300m,0.5)}
		\label{fig:dbscan_mal300}
	\end{subfigure}
	\begin{subfigure}{0.3\textwidth}
		\centering
		\includegraphics[width=\textwidth]{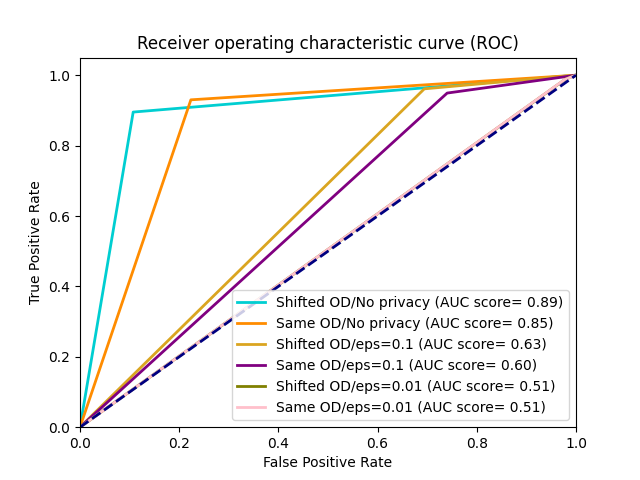}
		\caption{$(c,q)$=(500m,0.7)}
		\label{fig:dbscan_mal500}
	\end{subfigure}
	\begin{subfigure}{0.3\textwidth}
		\centering
		\includegraphics[width=\textwidth]{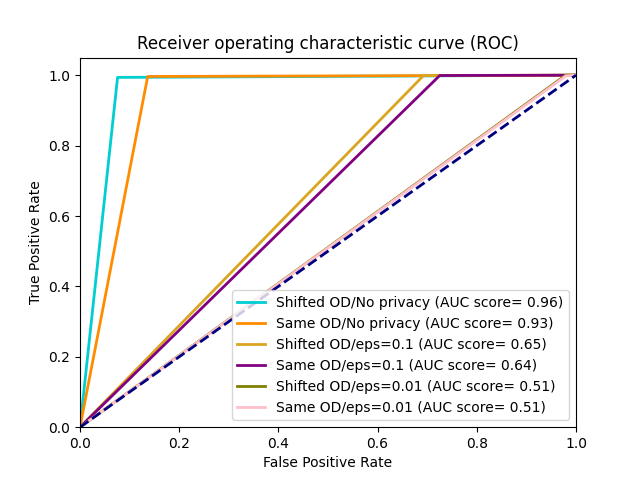}
		\caption{$(c,q)$=(700m,1.0)}
		\label{fig:dbscan_mal700}
	\end{subfigure}
	\caption{The influence of origin-destination locations on the classification accuracy of DBSCAN for different values of malicious intent.}
	\label{fig:dbscan_OD}
\end{figure*}

\begin{figure*}[tb!]
	\centering
	\begin{subfigure}{0.3\textwidth}
		\centering
		\includegraphics[width=\textwidth]{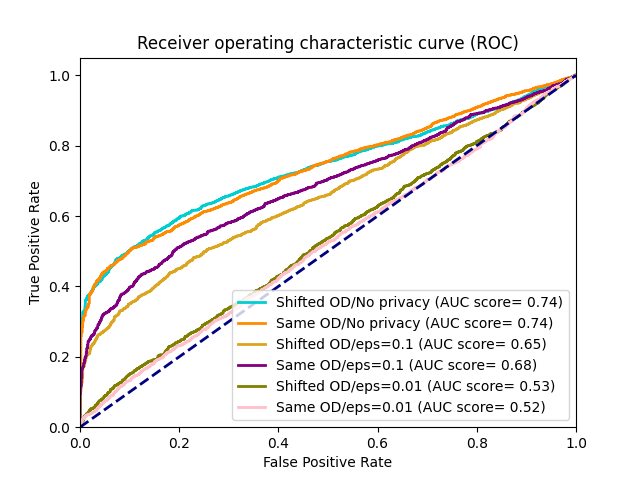}
		\caption{$(c,q)$=(300m,0.5)}
		\label{fig:rnn_od_mal300}
	\end{subfigure}
	\begin{subfigure}{0.3\textwidth}
		\centering
		\includegraphics[width=\textwidth]{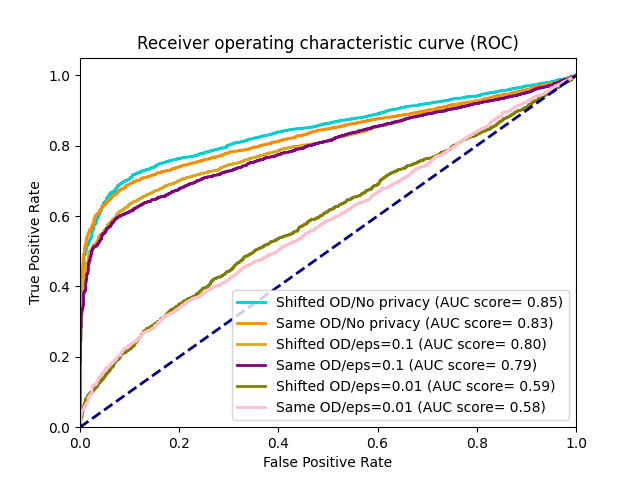}
		\caption{$(c,q)$=(500m,0.7)}
		\label{fig:rnn_od_mal500}
	\end{subfigure}
	\begin{subfigure}{0.3\textwidth}
		\centering
		\includegraphics[width=\textwidth]{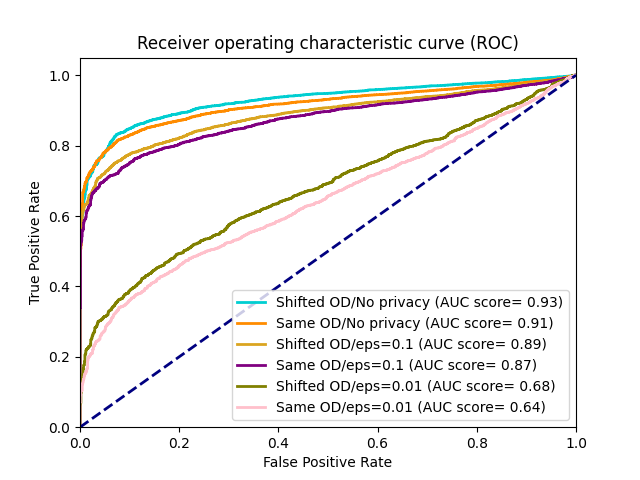}
		\caption{$(c,q)$=(700m,1.0)}
		\label{fig:rnn_od_mal700}
	\end{subfigure}
	\caption{The influence of origin-destination locations on the classification accuracy of RNN for different values of malicious intent.}
	\label{fig:rnn_OD}
\end{figure*}

In addition to privacy-preservation mechanisms, we also evaluate the influence of shifting O-D pairs of the generated anomalous trips on the two classifiers. This is because adversaries with the prior knowledge of possible origin and destination locations that normal drivers may visit can intentionally generate malicious trips which looks more plausible: each trip has an origin and a destination located in the same region as normal trips. Fig.~\ref{fig:dbscan_OD} and~\ref{fig:rnn_OD} illustrate the influence of O-D locations on the performance of DBSCAN and RNN respectively. The legend of ``shifted OD'' in both figures means that adversaries will randomly generate locations of origins and destinations when building anomalous trips, while the legend of ``same OD'' means that adversaries will carefully choose origins and destinations which are close to those O-D pairs visited by normal drivers based on prior knowledge.  

For DBSCAN, when no privacy perturbation is applied to the original vehicle trip data, the classifier achieves better performance in finding anomalies with shifted O-D pairs (cyan curves in Fig.~\ref{fig:dbscan_OD}) than trips that contains same O-D pairs (orange curves in Fig.~\ref{fig:dbscan_OD}), as shown in Fig.~\ref{fig:dbscan_mal300},~\ref{fig:dbscan_mal500}, and~\ref{fig:dbscan_mal700}. For example, when no privacy perturbation is applied to the trip data and malicious intent value is (300m,0.5), DBSCAN can achieve 0.82 AUC score in detecting anomalous trips with shifted O-D pairs, while the classifier only achieves 0.75 AUC score in detecting anomalous trips with same O-D pairs. However, the performance difference becomes smaller as more noise is added to perturb the trip data. For privacy perturbation with a noise level of $\epsilon$ = 0.1, only 0.02 difference is observed between shifted and same O-D pairs, as shown in Fig.~\ref{fig:dbscan_mal300}. 

For RNN, when no privacy perturbation is applied to the original vehicle trip data, the classifier achieves almost the same performance in finding anomalies between shifted O-D pairs (cyan curves in Fig.~\ref{fig:rnn_od_mal300}) and trips that contains same O-D pairs (orange curves in Fig.~\ref{fig:rnn_od_mal300}). 

These results are consistent with those we presented earlier in Figures~\ref{fig:twoclassifier_mal300m},~\ref{fig:twoclassifier_mal500m}, and~\ref{fig:twoclassifier_mal700m}, confirming that DBSCAN is more vulnerable than RNN to location perturbation for achieving differential privacy. The general remarks on the performance of DBSCAN and RNN when applying trip location privacy perturbation are reported in Table~\ref{tab:results}. 

\subsection{Recommended usage scenarios of the anomaly detectors}
Based on the results from the experimental evaluation presented in this paper, when models for anomaly detection have to be chosen, we recommend  engineers to take three aspects into account: customers' need for privacy preservation (i.ee., level of privacy); the size of the available vehicle trip data; and the required timing constraints of the specific mobility service provider.

Since DBSCAN is vulnerable to 2d noise, detectors based on deep neural networks are a more appropriate choice if there is the need of applying privacy perturbation to vehicle locations before data is sent from the edge devices to the cloud servers. 

Additionally, DBSCAN suffers from scalability issues due to the large number of computations needed for calculating trip similarities, which grows exponentially witht the number of trips. The scalability is not only a DBSCAN issue, but it is related to other unsupervised anomaly detectors that use the notion of trajectories similarity in clustering. On this respect, the RNN technique discussed in this paper, such as other deep learning methods, are much more efficient as they only rely on the difference (e.g., mean square error) between reported locations and the location reconstructed by the neural networks to determine if a given candidate trip is anomalous. 

Despite these limitations, clustering-based unsupervised learning algorithms are still widely used in finding trip anomalies or discovering interesting travel patterns for small trip datasets. Furthermore, DBSCAN can be directly applied to vehicle trips without pre-training,  while RNN requires large amount of trip data for training the neural networks. The experimental results presented here suggest that DBSCAN and RNN have similar performance while the number of vehicle trips required by DBSCAN is less than $10\%$ of those required by RNN.

\section{Conclusion}

While increasing location privacy in LBS could protect
users’ data in case of breaches, it could also lead
to security issues for users while using these systems, arising the
question of which aspect should be prioritized. In this paper we investigate the tradeoff between security and location privacy in the context of ride-hailing services, by examining the impact of  location data privacy-preservation  on the performance of the anomaly detectors.
For identifying anomalous trips, we used a Density-based Spatial Clustering of Applications with Noise (DBSCAN) and a Recurrent Neural Network (RNN) framework, while to guarantee location privacy, we perturbed the original trip data
by applying two dimensional Laplace noise, considering both the individual location (location-based perturbation), and the potential temporal correlations between locations within the same trajectory (trajectory-based perturbation). We also investigated the effect on the detectors of shifting O-D pairs.
The results clearly show that while the level of privacy increases, by increasing the perturbation noise, the capacity for the system to identify anomalies could decrease quite sensibly, especially when using the clustering.

The general insights from the experimental evaluation are: i) Unsupervised learning techniques building on deep neural networks are more robust to differential location privacy than unsupervised learning techniques based on clustering techniques; ii) The privacy and security tradeoff is different between neural networks and clustering approaches for vehicle routes with origin-destination pairs not present in the original datasets. However, the impact is less obvious for neural networks than for clustering; iii) While neural networks are more robust to differential privacy noise than clustering, clustering-based methods detectors are still good candidates for anomaly detection when the size of the dataset is small (e.g., in the magnitude of 10 to 100 thousand trips but less than one million). The reason is that clustering models can be applied directly without training, while neural networks in our experiments requires a dataset with more than one hundred million trips for training.


The case study provides a concrete example of the tradeoff between the utility and the risk of location data collected from real-world crowd sensing platforms~\cite{calacci2019tradeoff}. This enables more effective evidence-based policy-making in laws and regulations for location data privacy. On one hand, an LBS customer should be given the option of selecting the level of privacy protection (s)he prefers. On the other hand, the potential influence of deploying privacy-preservation techniques on mobility services should be well understood. In addition to requiring mobility service providers to communicate the potential security risks and let customers be aware of the collection of their location data, policy makers may provide more guidance on how companies may adapt to customers' different levels of privacy needs when deciding on their privacy protection solutions.

Our future work will further explore the tradeoff between cyber-security and privacy from both policy and technical perspectives. For policy-making, we will investigate visualization tools for effectively communicating complex mathematical concept in location privacy, such as $\epsilon$ for differential privacy, to customers, developers of mobile sensing platforms, and policy makers. For technological development, we will compare differential privacy with other location masking techniques and their influence on anomaly detectors.

\ifCLASSOPTIONcaptionsoff
  \newpage
\fi




\bibliographystyle{IEEEtran}
\bibliography{IEEEabrv.bib,root.bib}
%

%

\newpage

\begin{IEEEbiography}[{\includegraphics[width=1in,height=1.25in,clip,keepaspectratio]{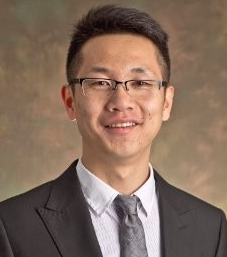}}]{Dajiang Suo} is a postdoctoral associate at MIT Auto-ID Lab. He obtained a Ph.D. in Mechanical Engineering from MIT in 2020. Suo holds a B.S. degree in mechatronics engineering, and S.M. degrees in Computer Science and Engineering Systems. His research interests include the trustworthiness and privacy preservation of multimodal vehicle data. Suo also worked on polarization-based sensing technologies for object recognition in autonomous vehicles. 

Before returning to school to pursue PhD degree, Suo was with the vehicle control and autonomous driving team at Ford Motor Company (Dearborn, MI), working on the safety and cyber-security of automated vehicles. He also serves on the Standing Committee on Enterprise, Systems, and Cyber Resilience (AMR40) at the Transportation Research Board.
\end{IEEEbiography}

\vskip -2\baselineskip plus -1fil

\begin{IEEEbiography}[{\includegraphics[width=1in,height=1.25in,clip,keepaspectratio]{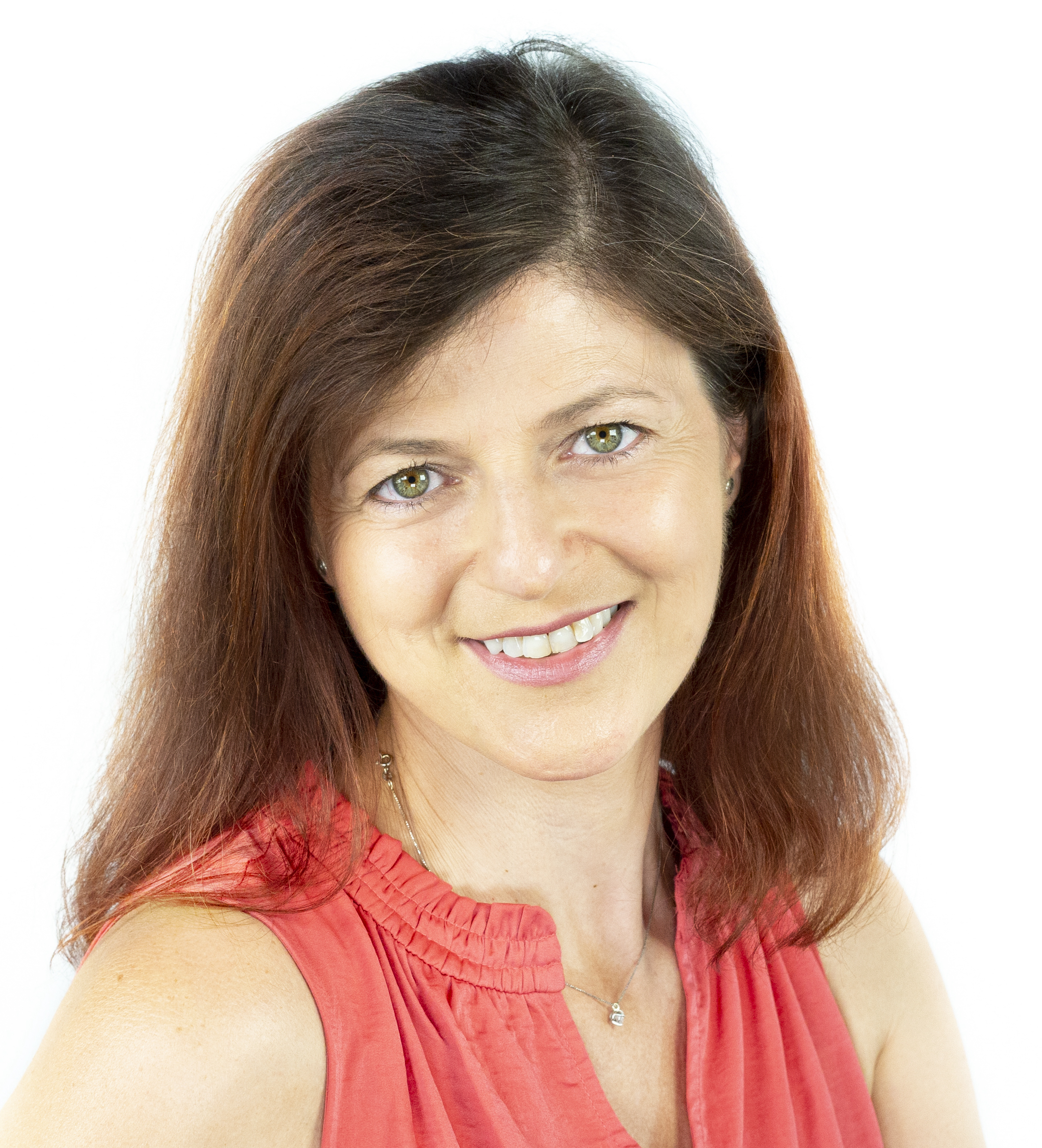}}]{M. Elena Renda} is a visiting Research Scientist within Department of Urban Studies and Planning at the Massachusetts Institute of Technology in Cambridge, US, and a Researcher at Istituto di Informatica e Telematica of the National Research Council (IIT-CNR) in Pisa, Italy. Dr. Renda holds a MS degree in Computer Science, and a Ph.D. Degree in Information Engineering. She has worked within the field of Information Retrieval, P2P, VANets, Mesh Networks, and algorithms for Bioinformatics. In the last decade Dr Renda has grown her interest towards Intelligent Transportation Systems, Smart Cities and Public Transportation. In particular she is working on defining, studying and exploiting new mobility paradigms (such as shared mobility) and evaluating their environmental, economic, and social effects. Privacy issues and their implications within the smart mobility realm is one of her most recent research interests.
\end{IEEEbiography}

\vskip -2\baselineskip plus -1fil

\begin{IEEEbiography}[{\includegraphics[width=1in,height=1.25in,clip,keepaspectratio]{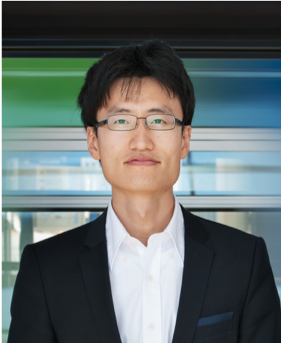}}]{Jinhua Zhao} is the Associate Professor of City and Transportation Planning at the Massachusetts Institute of Technology (MIT). Prof. Zhao brings behavioral science and transportation technology together to shape travel behavior, design mobility system, and reform urban policies. He develops methods to sense, predict, nudge, and regulate travel behavior and designs multimodal mobility systems that integrate automated and shared mobility with public transport. He sees transportation as a language to describe a person, characterize a city, and understand an institution and aims to establish the behavioral foundation for transportation systems and policies.
Prof. Zhao directs the JTL Urban Mobility Lab and Transit Lab at MIT and leads long-term research collaborations with major transportation authorities and operators worldwide, including London, Chicago, Hong Kong, and Singapore. He is the co-director of the Mobility Systems Center of the MIT Energy Initiative, and the director of the MIT Mobility Initiative.

\end{IEEEbiography}







\end{document}